\documentclass[10pt,twocolumn,letterpaper]{article}

\usepackage[pagenumbers]{cvpr} % To force page numbers, e.g. for an arXiv version
\definecolor{cvprblue}{rgb}{0.21,0.49,0.74}
\usepackage[pagebackref,breaklinks,colorlinks,allcolors=cvprblue]{hyperref}
\usepackage{courier}
\usepackage{tabularx}
\def\paperID{57} % *** Enter the Paper ID here
\def\confName{CVPR}
\def\confYear{2025}

\title{Synthetic Generation and Latent Projection Denoising of Rim Lesions in Multiple Sclerosis}

\author{\qquad Alexandra G. Roberts\\
{\qquad \tt\small agr78@cornell.edu}
\and  \qquad
Ha M. Luu\\
{\qquad \tt\small hal4025@med.cornell.edu}
\and
Mert \c{S}i\c{s}man\\
{\qquad \tt\small ms2893@cornell.edu}
\and \qquad
Alexey V. Dimov\\
{\qquad \tt\small ald2031@med.cornell.edu}
\and
Ceren Tozlu\\
{\tt\small cet2005@med.cornell.edu}
\and 
Ilhami Kovanlikaya\\
{\tt\small ilk2002@med.cornell.edu}
\and
Susan A. Gauthier\\
{\tt\small sag2015@med.cornell.edu}
\and
Thanh D. Nguyen\\
{\tt\small tdn2001@med.cornell.edu}
\and
Yi Wang\\
{\tt\small yiwang@med.cornell.edu}\\
}

\begin{document}
\maketitle
\begin{abstract}
Quantitative susceptibility maps from magnetic resonance images can provide both prognostic and diagnostic information in multiple sclerosis, a neurodegenerative disease characterized by the formation of  lesions in white matter brain tissue. In particular, susceptibility maps provide adequate contrast to distinguish between ``rim'' lesions, surrounded by deposited paramagnetic iron, and ``non-rim'' lesion types. These paramagnetic rim lesions (PRLs) are an emerging biomarker in multiple sclerosis. Much effort has been devoted to both detection and segmentation of such lesions to monitor longitudinal change. As paramagnetic rim lesions are rare, addressing this problem requires confronting the class imbalance between rim and non-rim lesions. We produce synthetic quantitative susceptibility maps of paramagnetic rim lesions and show that inclusion of such synthetic data improves classifier performance and provide a multi-channel extension to generate accompanying contrasts and probabilistic segmentation maps. We exploit the projection capability of our trained generative network to demonstrate a novel denoising approach that allows us to train on ambiguous rim cases and substantially increase the minority class. We show that both synthetic lesion synthesis and our proposed rim lesion label denoising method best approximate the unseen rim lesion distribution and improve detection in a clinically interpretable manner. We release our code and generated data at \texttt{\href{https://github.com/agr78/PRLx-GAN}{https://github.com/agr78/PRLx-GAN}} upon publication.
\vfill
\end{abstract}    
\section{Introduction}
\label{sec:intro}
\begin{figure}[t]
  \centering
  % \fbox{\rule{0pt}{2in} \rule{0.9\linewidth}{0pt}}
   \includegraphics[width=\linewidth]{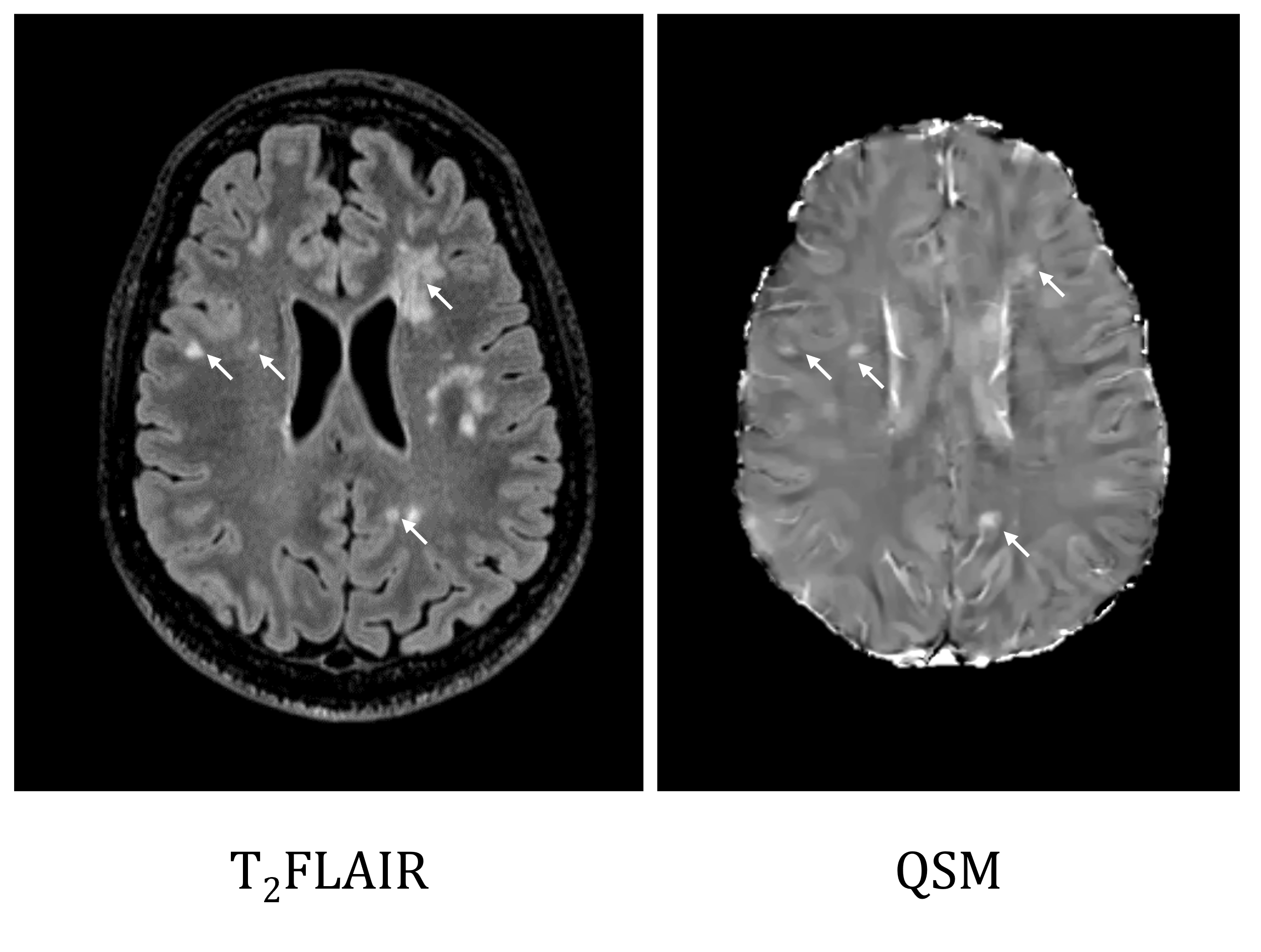}
   \caption{Example of MS patient with lesions depicted on qualitative $\mathrm{T_2FLAIR}$ and whole brain QSM. Lesions appearing on both contrasts are indicated by arrows.}
   \label{f1}
\end{figure}
\subsection{Background}
Multiple sclerosis (MS) is a debilitating neurodegenerative disease~\cite{Gmey} with a rising global presence~\cite{Cwal}. MS is the most common demyelinating disorder~\cite{Pwil} and there is currently no cure~\cite{Kgoh}, though a variety of disease-modifying therapies exist to slow symptom progression and improve quality of life~\cite{Drob}. 

The efficacy of these therapies is measured by the appearance of gadolinium-enhancing lesions in brain white matter as identified on magnetic resonance imaging (MRI)~\cite{Ched}. The importance of MRI in both MS diagnosis and prognosis~\cite{Ukau} motivates the development of automatic lesion detection and segmentation algorithms to address this growing clinical need~\cite{Rben}.

% \subsection*{Paper ID}
% Make sure that the Paper ID from the submission system is visible in the version submitted for review (replacing the ``*****'' you see in this document).
% If you are using the \LaTeX\ template, \textbf{make sure to update paper ID in the appropriate place in the tex file}.

% \subsection*{Mathematics}

% Please number all of your sections and displayed equations as in these examples:
% \begin{equation}
%   \mathbb{E}_{\mathbf{w},\mathbf{y}\sim\mathcal{N}(0,\mathbf{I})} \left[\left(\left|\left|\nabla_{\mathbf{w}}g(\mathbf{w})\cdot\mathbf{y}\right|\right|_2-a\right)^2\right]
%   \label{eq:important}
% \end{equation}

%-------------------------------------------------------------------------
\subsection{Imaging techniques}
Much progress on the understanding of MS has been gained through the qualitative $\mathrm{T_1w}$ and $\mathrm{T_2FLAIR}$ MRI contrasts~\cite{Mfil}. However, longitudinal studies increasingly rely on quantitative susceptibility maps (QSM) as robust biomarkers in MS disease progression~\cite{Szha,Cfis,Wche,Cvoo}. Of critical importance is the visualization of paramagnetic rim lesions linked to symptom severity~\cite{Ator}. 

Due to the iron depositions surrounding the rim of these lesions~\cite{Ahof}, susceptibility contrasts like QSM are required to differentiate between lesion subtypes~\cite{Whua}. Co-registered $\mathrm{T_2FLAIR}$ and whole brain QSM~\cite{Arob} are shown in an example MS patient in Figure \ref{f1}, with lesions indicated by the white arrows. Iron is involved in a variety of neuroinflammatory diseases and deposition often increases with inflammatory response~\cite{Rwar}. Activated immune cell microglia at the edge of the lesion are the primary source of iron, generating contrast between the paramagnetic rim and diamagnetic lesion core~\cite{Jnam}. 

\begin{figure}[t]
  \centering
  % \fbox{\rule{0pt}{2in} \rule{0.9\linewidth}{0pt}}
   \includegraphics[width=\linewidth]{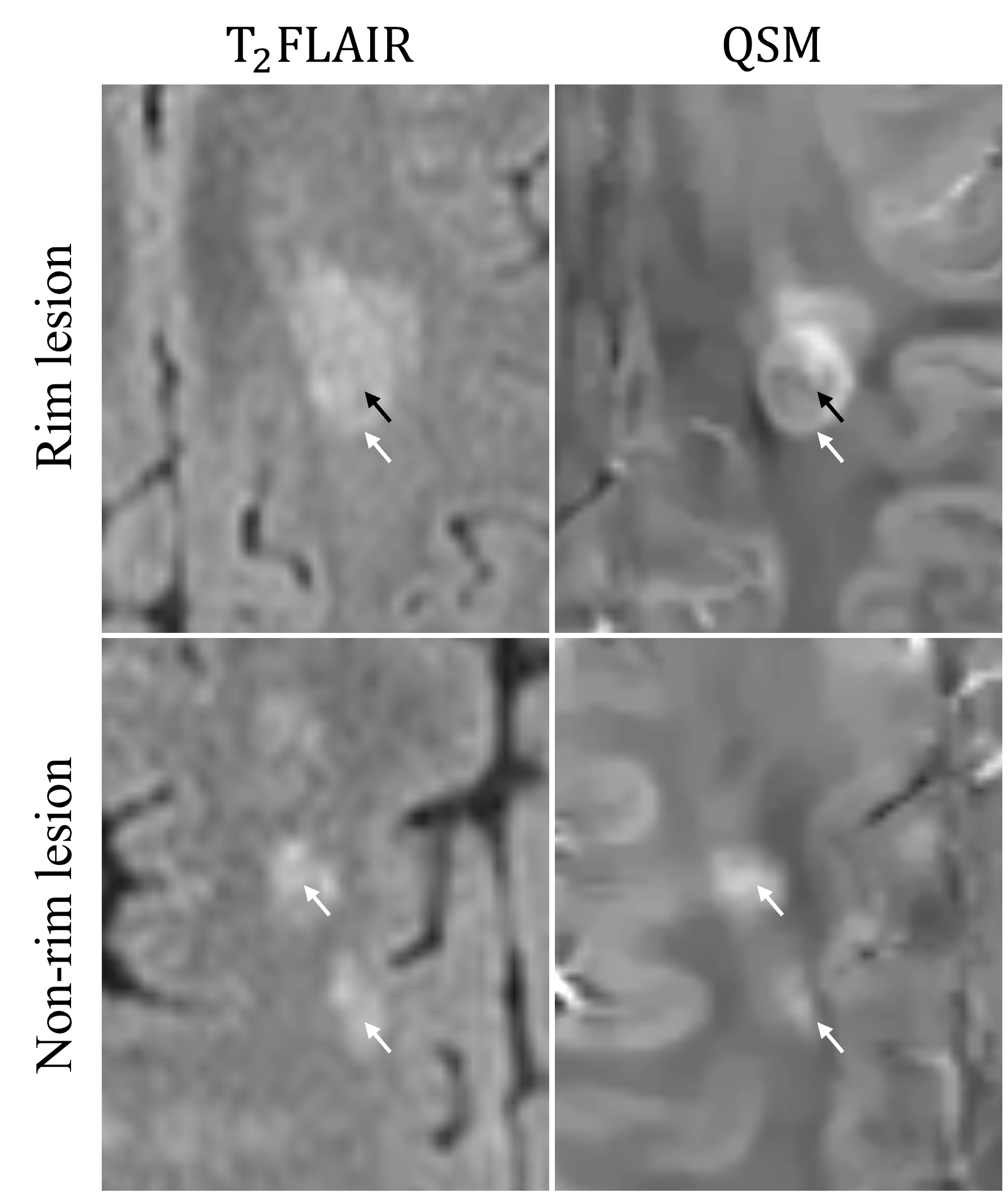}
   \caption{Example of rim (top) and non-rim (bottom) lesion visualization on $\mathrm{T_2FLAIR}$ (left) and QSM (right). Note that rim lesions can only be differentiated from non-rim lesions on QSM. The rim lesion is composed of a hypointense core (black arrow) and hyperintense rim (white arrow), contrast to a hyperintense non-rim lesion.}
   \label{f2}
\end{figure}

As QSM directly measures small changes in the applied magnetic field arising from paramagnetic or diamagnetic tissue content, it is a reliable method to detect rim lesions~\cite{Jree}. Figure \ref{f2} illustrates the need for QSM to enable rim identification as compared to $\mathrm{T_2FLAIR}$. 

%-------------------------------------------------------------------------

\subsection{Lesion class imbalance}
Though paramagnetic rim lesions differentiate between MS and other neurodegenerative disorders with high specificity~\cite{Pmag}, only $10\%$ of all MS lesions are estimated to be rim lesions~\cite{Kwon}, introducing a class imbalance problem for detection and segmentation algorithms. 

Despite consensus~\cite{Fbag} on rim lesion characteristics, ambiguous cases remain when expert readers disagree on these criteria. It is desirable to ``denoise'' such lesions in order to make use of all possible rim lesion data.

\subsection{Contributions}
We present a novel sample denoising approach based on generated synthetic quantitative susceptibility maps of paramagnetic rim lesions. We exploit the projection capability~\cite{Acre} in our trained generative network to "denoise" ambiguous samples by recovering unambiguous synthesized samples. We evaluate the effect of training with said data in the rim classification problem and demonstrate improvement in trained detectors when denoised synthetic data is included. This approach enables us to train with noisy or ambiguous labels by recovering their synthetic analog from the trained generator network. To our knowledge, our proposed method is the first to allow ambiguous or contested rim lesion labels to augment the rare, unambiguous rim lesion class during rim detection. Finally, we provide a multi-contrast extension to enable generation of susceptibility maps, $\mathrm{T_2FLAIR}$ images, and probabilistic rim lesion segmentations, broadening the utility of our framework.

% \begin{figure*}
%   \centering
%   \begin{subfigure}{0.68\linewidth}
%     \fbox{\rule{0pt}{2in} \rule{.9\linewidth}{0pt}}
%     \caption{An example of a subfigure.}
%     \label{fig:short-a}
%   \end{subfigure}
%   \hfill
%   \begin{subfigure}{0.28\linewidth}
%     \fbox{\rule{0pt}{2in} \rule{.9\linewidth}{0pt}}
%     \caption{Another example of a subfigure.}
%     \label{fig:short-b}
%   \end{subfigure}
%   \caption{Example of a short caption, which should be centered.}
%   \label{fig:short}
% \end{figure*}

\section{Related Works}
\label{sec:formatting}
\subsection{Synthetic MRI}
Synthetic MRI data ranges from classical representations generated via signal models~\cite{Mcal,Jand} to more recent deep learning approaches aiming to estimate mappings over synthetic data that are applicable to $in \ vivo$ data~\cite{Emoy,Kgop} and across various contrasts and resolutions~\cite{Jigl}. Given pretrained models and approximate source and target distributions~\cite{jwil}, transfer learning approaches can address the challenges of gathering sufficient MRI data~\cite{weli,jval,smat}. 

Another subset of solutions focus on generating synthetic data for small datasets~\cite{Vtha} and label scarcity arising from new~\cite{Awah} or lethal~\cite{Bahm} pathologies. Rim segmentation efforts have addressed the class imbalance problem by oversampling the minority class in the latent space~\cite{Hzha,Ddab}, included in our comparison. 

Synthetic lesions in multiple sclerosis have been generated~\cite{Msal} using a variational autoencoder on qualitative $\mathrm{T_2FLAIR}$. Like our work, this model aims to generate synthetic MS lesions, differing in that the whole brain learned mapping is between healthy cases and MS patients on qualitative $\mathrm{T_2FLAIR}$. 

We train a generative adversarial network~\cite{Igoo} (GAN) with the goal of learning the mapping from random noise initializing the generator model to synthetic, quantitative, paramagnetic rim lesions. We do this using susceptibility images as rim lesions are differentiable from non-rim lesions only on QSM. 
\subsection{Latent projection denoising}
From our choice in architecture arises an opportunity to recover ambiguous or ``noisy'' rim lesions by projecting their latent vectors into the latent space of a pretrained GAN synthesizing only unambiguous rim lesions, closely related to GAN inversion~\cite{Wxia}. Namely, these ambiguous cases are lesions where expert raters disagree on the label. The presence of label noise has been addressed by conditional GANs~\cite{Mmir,Kthe} (cGAN), which are trained on both majority and minority class labels. 

Given the data imbalance between rim and non-rim lesions, we focus on the minority class rather than the majority class and we show its inclusion nearly doubles the required training time. We seek to make use of ambiguous, noisy real rim lesions by training the generator only the unambiguous rim lesion minority class. We use the projection into the learned latent space from unambiguous (or noiseless) rim lesions to recover denoised lesions. Other works related to this effort include modeling label noise as a latent space shift~\cite{Wehua}, a technique applied to correct classifications rather than augment data. 

Also related is estimation of the noise transition matrix~\cite{Hbae} to calibrate classifiers trained on noisy labels, which requires some understanding of the noise distribution. Perhaps most relevant is the use of GAN inversion for under-sampled MRI reconstruction~\cite{Vkel}, which deals with the presence of instrument noise rather than mitigation of more subjective label noise.
\section{Method}
\subsection{Dataset}
A group of $256$ MS subjects (mean age, $46.2 \pm 11.8 \ years$, $79$ men $(30.8\%)$, $177$ women $(69.2\%)$) were imaged on a $3T$ Magnetom Skyra scanner. An axial $3D$ multi-echo GRE sequence was used to acquire phase data for QSM with $FOV=24.0 \ cm$, $\frac{TE_1}{dTE} = 6.28/4.06 \ ms$,  $8$ echos, $T_R=40 \ ms$, voxel size $0.4 \times 0.4 \times 1 \ mm^3$. 

QSM was reconstructed  using referenced morphology-enabled  dipole  inversion (MEDI+0)~\cite{Zliu,Tliu}. The dataset was prepared as follows~\cite{Hluu}. Each lesion on QSM was cropped to an image patch of $64 \times 64 \times 24$ voxels. Two  expert  readers independently created rim lesion ground truth labels according to the recent consensus statement~\cite{Fbag}. 

Lesions were classified as ``rim'' only if both readers agreed on their paramagnetic rim lesion status, otherwise they were classified as ``non-rim''. Hyperintense rim areas of each identified lesion were manually traced and checked by the same two readers. Critical to our denoising approach is the definition of an \textbf{ambiguous rim lesion}, where only one of the two expert raters described the lesion as a paramagnetic rim lesion, and the label is noisy.

\subsection{Radiologist assessment}
An expert radiologist guided by the recent rim lesion consensus~\cite{Fbag} reviewed $110$ uncurated example slices of real and synthetic paramagnetic rim lesions. In two separate experiments, the lesions were classified as ``real'' or ``synthetic'' and categorized as ``rim'' or ``non-rim''. 
\subsection{Classifier network}
To evaluate the improvement in classification with added synthetic data, a simple convolutional neural network classifier for binary classification was implemented~\cite{Ylec}. The network consisted of $6$ convolutional layer units including pooling and batch normalization operations, followed by a rectified linear unit (ReLU) activation function. The classifier was trained for $25$ epochs with the Adam optimizer with learning rate $10^{-3}$ using the cross-entropy loss function.
\subsection{Generative network}
\begin{figure}[t]
  \centering
  % \fbox{\rule{0pt}{2in} \rule{0.9\linewidth}{0pt}}
   \includegraphics[width=\linewidth]{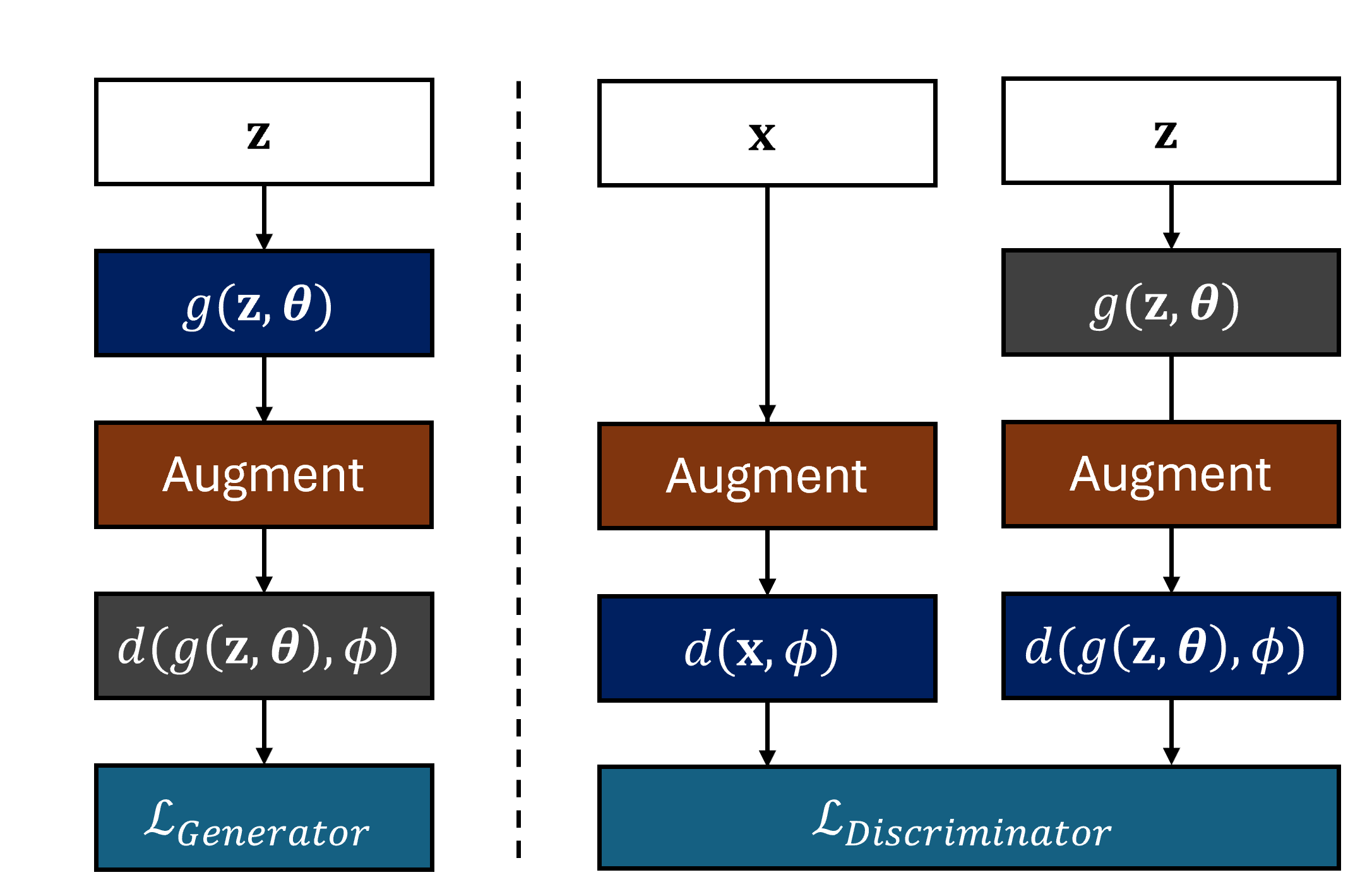}
   \caption{Schematic of the GAN with adaptive discriminator augmentation blocks used to generate synthetic rim lesion data from limited datasets. The architecture consists of generative and discriminative networks $g(\mathbf{z},\mathbf{\theta})$ and $d(.,\phi)$ and receives inputs real images $\mathbf{x}$ and latent variable $\mathbf{z} \sim \mathcal{N}(0,1)$. Adapted from~\cite{Tker}.}
   \label{f3}
\end{figure}

StyleGAN2 with adaptive discriminator augmentation (StyleGAN2-ADA, shortened in this work for brevity to ``ADA-GAN'')~\cite{Tker} was trained via Frechet Inception Distance~\cite{Mheu} (FID) minimization with the Adam optimizer (learning rate $2.5\times10^{-3}$, first and second moment decay rates $\beta_1=0.9$, $\beta_2=0.99$, respectively) using overfitting heuristic $r_t=0.6$. The training dataset contained 200 rim lesions and 400 non-rim lesions. For testing, $60$ rim lesions and $120$ non-rim lesions were withheld from both the generator and classifier. 
Training required $32$ hours with $8$ NVIDIA GeForce RTX $2080$ graphics cards to generate $25,000$ synthetic rim lesions. An overview of the architecture is given in Figure \ref{f3}.
\vfill\null
\subsection{Denoising approach}
\begin{figure}[t]
  \centering
  % \fbox{\rule{0pt}{2in} \rule{0.9\linewidth}{0pt}}
   \includegraphics[width=\linewidth]{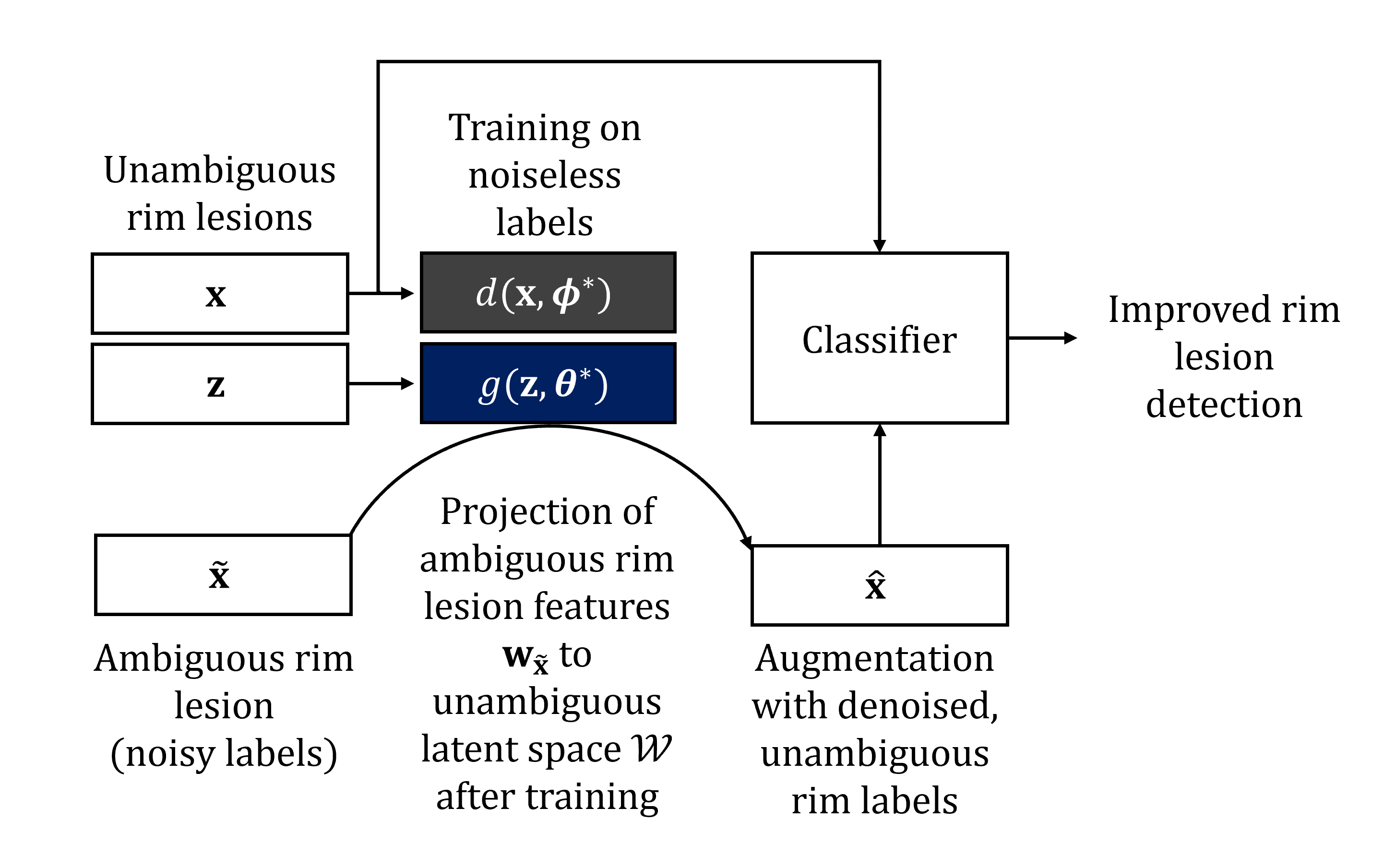}
   \caption{Outline of denoising approach, beginning with generative model training on unambiguous rim lesion $\mathbf{x}$. After training, features of ambiguous rim lesions (with noisy labels) are extracted to latent variable $\mathbf{w}_{\tilde{\mathbf{x}}}$ and projected onto the unambiguous latent space $\mathcal{W}$. Finally, the real rim lesion training data is augmented with these "denoised" rim lesions and the classifier performance is seen to improve.} 
   \label{f4}
\end{figure}

We exploit the projection capability of the chosen network~\cite{Tker1} to create synthetic denoised samples from ambiguous or noisy samples $\mathbf{\tilde{x}}$. We note that this particular application is not aiming to remove instrument or random noise within the image, but noise in the labeling process that results in ambiguous lesions where raters disagree. As such, the ambiguous rim lesion image is transformed to be unambiguous and the label is ``denoised''. Denoised samples $\hat{\mathbf{x}}=s(\mathbf{w}^*,\theta^*_s)$ are computed from the synthesis network $s$ module in the generator network $g$. The latent projection $\mathbf{w}^*$ is the ``closest'' intermediate latent space vector to noisy image $\mathbf{\tilde{x}}$ features, $\mathbf{w}_\mathbf{\tilde{x}}$, supported by trained generator $g$, described in Figure \ref{f4}. This optimal latent space vector $\mathbf{w}^*$ is obtained as in~\cite{Tker2}
\begin{equation}\label{eq1}
  \mathbf{w}^*=\mathrm{argmin}_{\mathbf{w}} L_{P}(\mathbf{\tilde{x}},s(\mathbf{w},\mathbf{\theta}_s^*))+\alpha\sum_{i,j}L_N(n_i,n_j)
\end{equation}
Where $L_P$ is the perceptual loss~\cite{Rzha} over extracted features~\cite{Ksim} at each layer and $L_{N}$ is the noise regularization term (scaled by parameter $\alpha$) from added noise vector $\mathbf{n}$ at original resolution $i$ and downsampled resolution $j$, $n_i\sim\mathcal{N}(0,I)$, and $n_j$. 

\begin{figure}[t]
  \centering
  % \fbox{\rule{0pt}{2in} \rule{0.9\linewidth}{0pt}}
   \includegraphics[width=\linewidth]{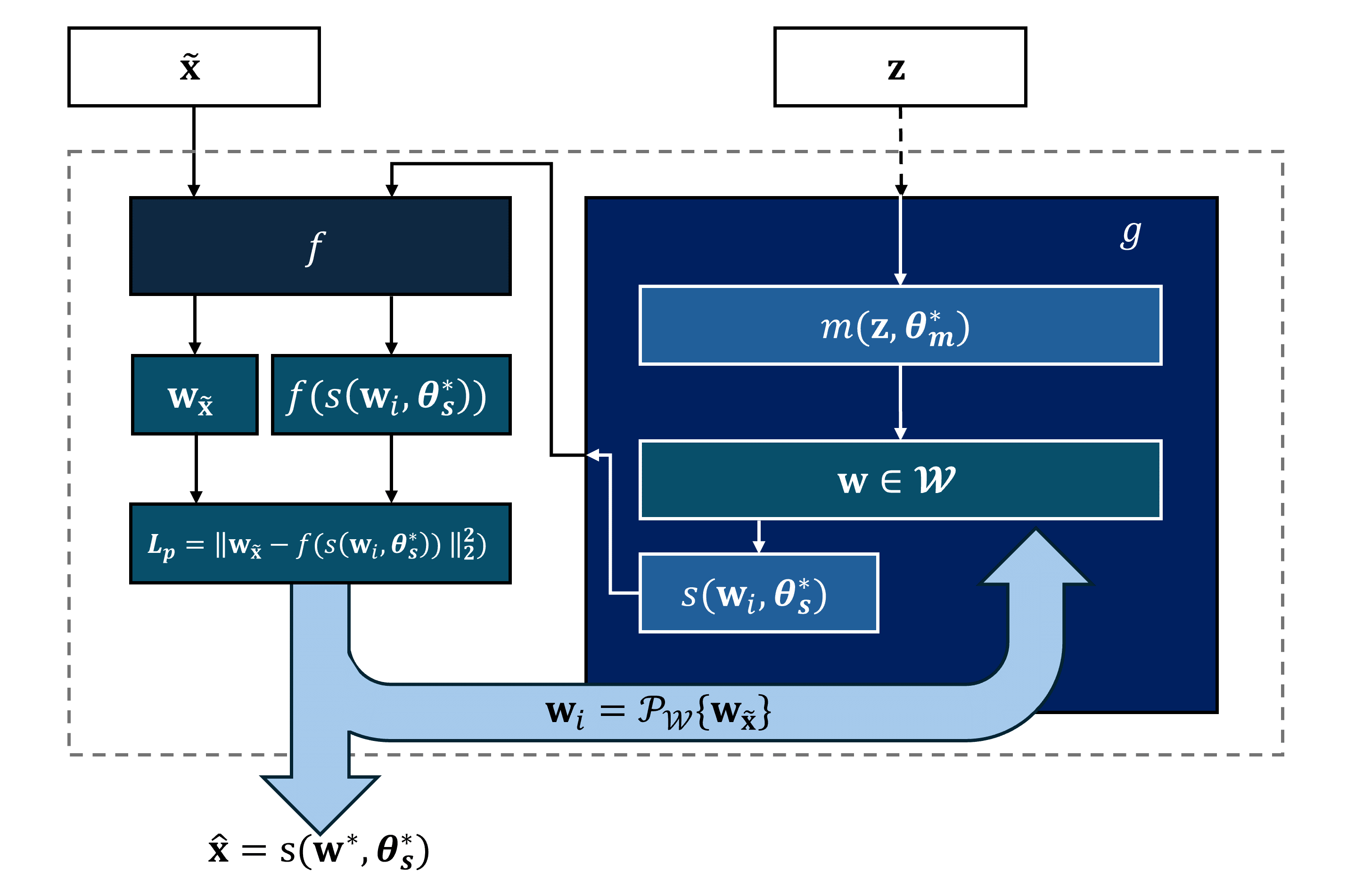}
   \caption{A simplified representation (adapted from~\cite{Tker1}) of the mapping $m$ and synthesis $s$ modules used to enable denoising via the trained generative network $g$. Given an ambiguous rim $\tilde{\mathbf{x}}$ lesion, denoising occurs via latent projection $\mathcal{P}$ after feature extraction by pretrained network $f$. The denoised projection $\mathbf{w}^*$ results from minimizing the perceptual loss $L_2$ term ($L_P$) with added noise regularization (Equation \ref{eq1}). After training with the architecture in Figure \ref{f3}, the synthesis module $s$ decodes the denoised projection $\hat{\mathbf{x}}$ corresponding to the "closest" unambiguous rim lesion in the latent space $\mathcal{W}$, enabling the augmentation in Figure \ref{f4}.} 
   \label{f5}
\end{figure}

Equation \ref{eq1} is solved using the Adam optimizer over $1000$ iterations with regularization $\alpha=10^5$, first and second moment decay rates $\beta_1=0.9$, $\beta_2=0.99$, and a scheduled learning rate initialized at $10^{-1}$. Rather than augmenting the dataset with synthetic rim lesions $\mathbf{x}'$ mapped from random noise, we augment with the denoised projections $\hat{\mathbf{x}}$ of noisy samples $\mathbf{\tilde{x}}$, related by $\hat{\mathbf{x}}=s(\mathbf{w}^*,\mathbf{\theta}_s^*)$ from the generator $g$ with weights $\mathbf{\theta^*}$. An overview is depicted in Figure \ref{f5}. We term this augmentation to increase the minority sample class via latent projection denoising as ``ADA-GAN-LD''.

\subsection{Ambiguous rim lesion denoising}
We apply the aforementioned denoising projection in the latent space of ADA-GAN to transform 100 ambiguous rim lesions - after training only on unambiguous rim lesions. We compare the resulting ADA-GAN-LD dataset to a number of other possible augmentations described in later sections. We also combine synthetic data and denoised rim lesions in an augmentation referred to as ``ADA-GAN+LD.


\subsection{Conditional generative network}
The standard network was retrained with a conditional GAN with ``rim'' and ``non-rim'' labels. The training dataset above was expanded to include $400$ non-rim lesions and required $48$ hours of computing time with $8$ NVIDIA GeForce RTX $2080$ graphics cards to generate $25,000$ synthetic rim and non-rim lesions.

\subsection{Augmentation comparison}
The generative model, ADA-GAN, was compared to simple affine transformation augmentation (``Affine'') and the learned synthetic minority oversampling technique, ``DeepSMOTE''~\cite{Ddab} by training separate classifier networks on each augmentation. An ablation study was also performed to find the optimal combination of synthetic, conditioned synthetic and/or denoised data. The FID was computed between each augmented training dataset distribution and the unseen test rim lesion distribution.

\subsection{Clinical interpretation}
To improve detection explainability, we produce class activation maps~\cite{Bzho,Rram} (CAMs) from the unaugmented classifier and the classifier augmented with our proposed denoising method. 
% \subsection{Projection into 3D}

\subsection{Multi-contrast generalization}
As paramagnetic rim lesion detection methods often rely on both susceptibility contrasts and $\mathrm{T_2FLAIR}$~\cite{Clou}, we extend the generative model and denoising approach with a contrast dimension comprised of susceptibility, rim mask, and $\mathrm{T_2FLAIR}$ channels. 

In particular, the inclusion of a rim lesion mask or probabilistic map allows the synthetic and denoised data to be used in training segmentation algorithms. The standard network was retrained with this extension and required $34$ hours of computing time with 8 NVIDIA GeForce RTX $2080$ graphics cards to generate $25,000$ multi-contrast maps.
\section{Results}
\subsection{Lesion distribution}
From the cohort, a total of $260$ rim lesions ($3.3\%$) and $7720$ non-rim lesions ($96.7\%$) were identified. Another 177 lesions were identified as ``rim'' by merely one of the two expert readers. Cohen's kappa agreement between the two readers is $0.73$ (substantial agreement). Out of $256$ patients, $92$ ($35.9\%$) had at least one rim lesion - $35$ ($13.7\%$) had $1$ rim lesion, 18 ($7\%$) had $2$ rim lesions, and $36$ ($14.1\%$) had from $3$ to $12$ rim lesions. The ambiguous rim lesions were defined as the $177$ paramagnetic rim lesions identified where one of the two expert raters classified the lesion as a rim lesion.

\subsection{Radiologist grading}
From $55$ real and $55$ synthetic lesions, an expert radiologist, estimated nearly half ($0.4$) of the synthetic rim lesions to be true rim lesions as opposed to about one third ($0.29$) of the real rim lesions. In a separate experiment, nearly a third ($0.31$) of the uncurated synthetic lesions were estimated to be real images alongside just over half ($0.55$) of the real rim lesions. We remark that the overall fraction of rim lesions identified is a result of grading lesions over a single slice rather than the entire $3D$ volume.

\begin{figure}[t]
  \centering
  % \fbox{\rule{0pt}{2in} \rule{\linewidth}{0pt}}
   \includegraphics[width=\linewidth]{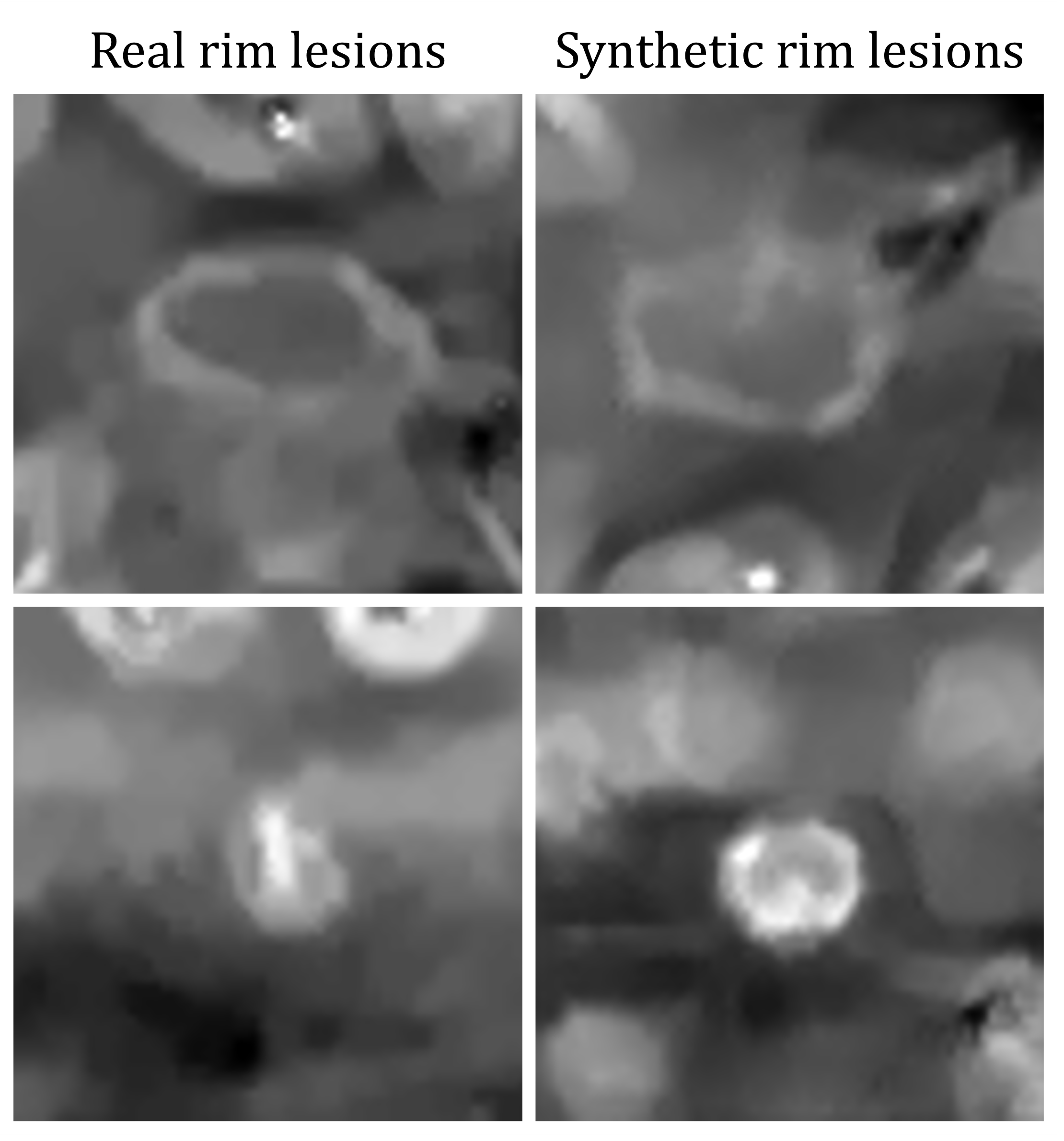}
   \caption{Typical real rim lesions compared to examples of synthetic rim lesions classified as real rim lesions.}
   \label{f6}
\end{figure}

We interpret these findings to suggest that convincing synthetic rims capture the less ambiguous features in rim definition, leading to a higher fraction of synthetic cases being classified as rim lesions. An example of an realistic synthetic rim (identified as real by an expert radiologist) and a similar real rim lesion is given in Figure \ref{f6}.
\begin{table}
  \centering
  \setlength{\tabcolsep}{9pt} 
  \begin{tabular}{@{}lcc@{}}
    \toprule
    Dataset & Rim lesion fraction & Real image fraction\\ 
    \midrule
    Real rims & 0.31 & 0.55 \\
    Synthetic rims & 0.4 & 0.29\\
    \bottomrule
  \end{tabular}
  \caption{Our expert radiologist estimated nearly half ($0.4$) of the synthetic rim lesions to be a true rim lesion as opposed to about one third ($0.31$) of the real rim lesions. In a separate experiment, and nearly a third ($0.29$) of the uncurated synthetic lesions were estimated to be real images alongside $0.55$ of the real rim lesions.}
  \label{tab:1}
\end{table}
\begin{table}
  \centering
  \setlength{\tabcolsep}{12pt} 
  \begin{tabular}{@{}lc@{}}
    \toprule
    Augmentation & FID from test rim lesions \\ 
    \midrule
    Real non-rim only & 46.35 \\
    DeepSMOTE rims & 36.48 \\
    Real rim only & 34.49 \\
    ADA-GAN rims & 34.36 \\
    ADA-GAN+LD rims (ours) & 34.24 \\
    \textbf{ADA-GAN-LD rims (ours)} & \textbf{34.17} \\
    \bottomrule
  \end{tabular}
  \caption{Computed FID between each training dataset distribution after augmentation and the unseen test rim lesion distribution. As expected, the majority non-rim lesion distribution is the furthest from the minority test rim lesion distribution. Real rim lesions, DeepSMOTE augmentation, and ADA-GAN augmentation all bring the training dataset of rim lesions closer to the unseen test rim lesion distribution, with our denoising approach minimizing the FID.}
 \label{tab:2}
\end{table}

\subsection{Rim lesion denoising}
\begin{figure}[t]
  \centering
  % \fbox{\rule{0pt}{2in} \rule{0.9\linewidth}{0pt}}
   \includegraphics[width=\linewidth]{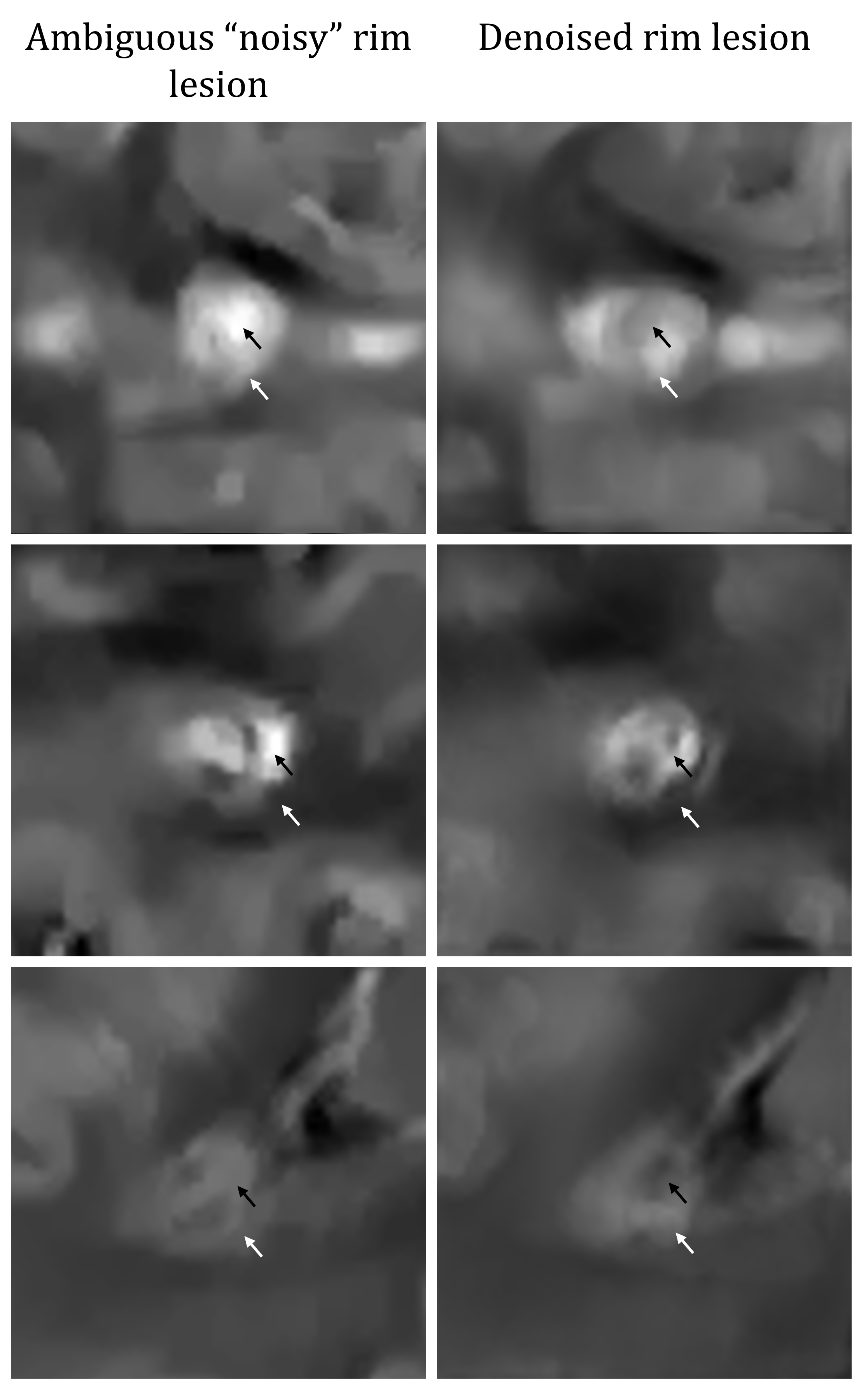}
   \caption{An ambiguous, ``noisy'' rim lesion $\mathbf{x}$ and the denoised projection $\hat{\mathbf{x}}$. Note the hypointensity (black arrows) at the denoised rim core not present in the original ambiguous rim lesion. Hyperintense rims are annotated with white arrows.}
   \label{f7}
\end{figure}

Our denoising method allowed us to recover $177$ additional rim lesions to expand the minority class from $260$ to $437$ cases. This augmentation increased our minority class labels by $68\%$. We projected $100$ noisy rim lesions $\tilde{\mathbf{x}}$ into the latent space $\mathcal{W}$ of the generator $g$ trained on only unambiguous rim lesions $\mathbf{x}$. We augment the classifier rim lesion training data with these denoised rim lesion samples and see optimal performance. Denoised example rim lesions $\hat{\mathbf{x}}$ are given in Figure \ref{f7}. 

We notice clinically interpretable changes to the denoised image $\hat{\mathbf{x}}$ when decoded from the unambiguous latent space by the synthesizer module $s$, such as defined hypointense lesion cores and the removal of obscuring artifacts.

\subsection{Classifier performance}
Evaluating the FID in Table \ref{tab:2} reveals the intuitive result that the majority non-rim lesion distribution is the furthest from the minority test rim lesion distribution. Real rim lesions without augmentation, DeepSMOTE augmentation, and ADA-GAN augmentation draw the training distribution of rim lesions closer to the unseen test rim lesion distribution, with our denoising augmentation approach minimizing the FID.
Augmenting the training data with $100$ synthetic rim lesions (``ADA-GAN'', ``DeepSMOTE'') improved the classifier performance as seen in Table \ref{tab:3}. In particular, accuracy and sensitivity were increased while the precision remained comparable to the dataset with no augmentation during training (``None''). Note that augmenting by simple random affine transformations (``Affine'') to increase the number of rim lesions (and replace an equal number of non-rim lesions) slightly improves classifier accuracy. We include our denoising approach (``ADA-GAN-LD'') to transform ``ambiguous rims'', which degrade classifier performance, into the unambiguous rim latent space, resulting in optimal classifier results. 

\begin{figure}[t]
  \centering
  % \fbox{\rule{0pt}{2in} \rule{0.9\linewidth}{0pt}}
   \includegraphics[width=\linewidth]{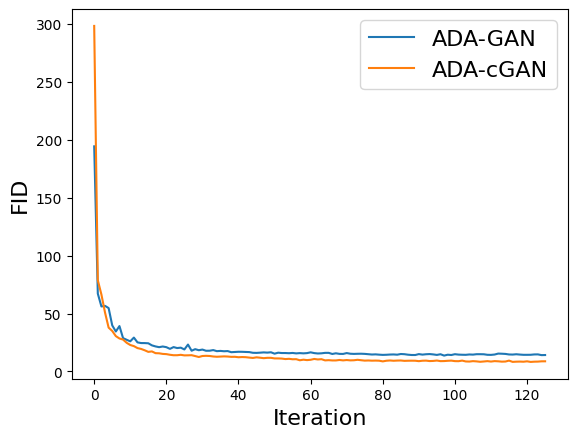}
   \caption{Conditional and standard GAN training curves. Inclusion of the non-rim label allows the training loss to converge to a slightly smaller FID.}
   \label{f8}
\end{figure}

\begin{table}
  \centering
  \setlength{\tabcolsep}{5pt} 
  \begin{tabular}{@{}lccc@{}}
    \toprule
    Augmentation & Accuracy & Precision & Sensitivity \\
    \midrule
    Ambiguous rims & 0.78 & 0.93 & 0.83 \\
    None & 0.79 & 0.93 & 0.85 \\
    Affine & 0.83 & 0.92 & 0.83 \\
    DeepSMOTE & 0.84 &  0.92 & 0.91 \\
    \textbf{ADA-GAN-LD (ours)} & \textbf{0.87} & 0.91 & \textbf{0.95} \\
    \bottomrule
  \end{tabular}
  \caption{Classifier performance using various training datasets. Including the noisy label ambiguous rim in the minority rim class results in a slight decrease  in performance. All augmentation strategies improve test results, with our denoising approach improving accuracy and sensitivity and offering comparable precision.}
  \label{tab:3}
\end{table}

\subsection{Ablation study}
We compare the standard ADA-GAN architecture to its conditional variant (ADA-cGAN) by comparing training loss curves to determine the effect of training the generator on both rim and non-rim lesions. We note that the conditional generative model, ADA-cGAN, converges to a slightly lower FID ($8.79$ compared to $14.20$) during training as seen in Figure \ref{f8}. We also compare our denoising method (ADA-GAN-LD) and combine the denoised and synthetic data (ADA-GAN+LD). We find that all variations of improve classifier performance, particularly our proposed projection denoising augmentation. Including the denoised rim lesions improves classifier accuracy and sensitivity and offers comparable precision as seen in Table \ref{tab:4}.
\begin{table}
  \centering
  \setlength{\tabcolsep}{5pt} 
  \begin{tabular}{@{}lccc@{}}
    \toprule
    Augmentation & Accuracy & Precision & Sensitivity \\
    \midrule
    None & 0.79 & 0.93 & 0.85 \\
    ADA-GAN & 0.85 & 0.91 & 0.93 \\
    ADA-GAN+LD (ours) & 0.85 & 0.92 & 0.92 \\
    ADA-cGAN & 0.86 & 0.91 & 0.93  \\
    \textbf{ADA-GAN-LD (ours)} & \textbf{0.87} & 0.91 & \textbf{0.95} \\
    \bottomrule
  \end{tabular}
  \caption{Classifier performance during ADA-GAN ablation study. We compare the addition of synthetic data, synthetic and denoised data, conditional synthetic data, and our proposed denoising approach as possible classifier augmentations. All augmentations improve classifier performance, particularly the conditional augmentation and the inclusion of denoised ambiguous rim lesions.}
  \label{tab:4}
\end{table}

\newpage

\subsection{Class activation maps} 
\begin{figure}[t]
  \centering
  % \fbox{\rule{0pt}{2in} \rule{\linewidth}{0pt}}
   \includegraphics[width=\linewidth]{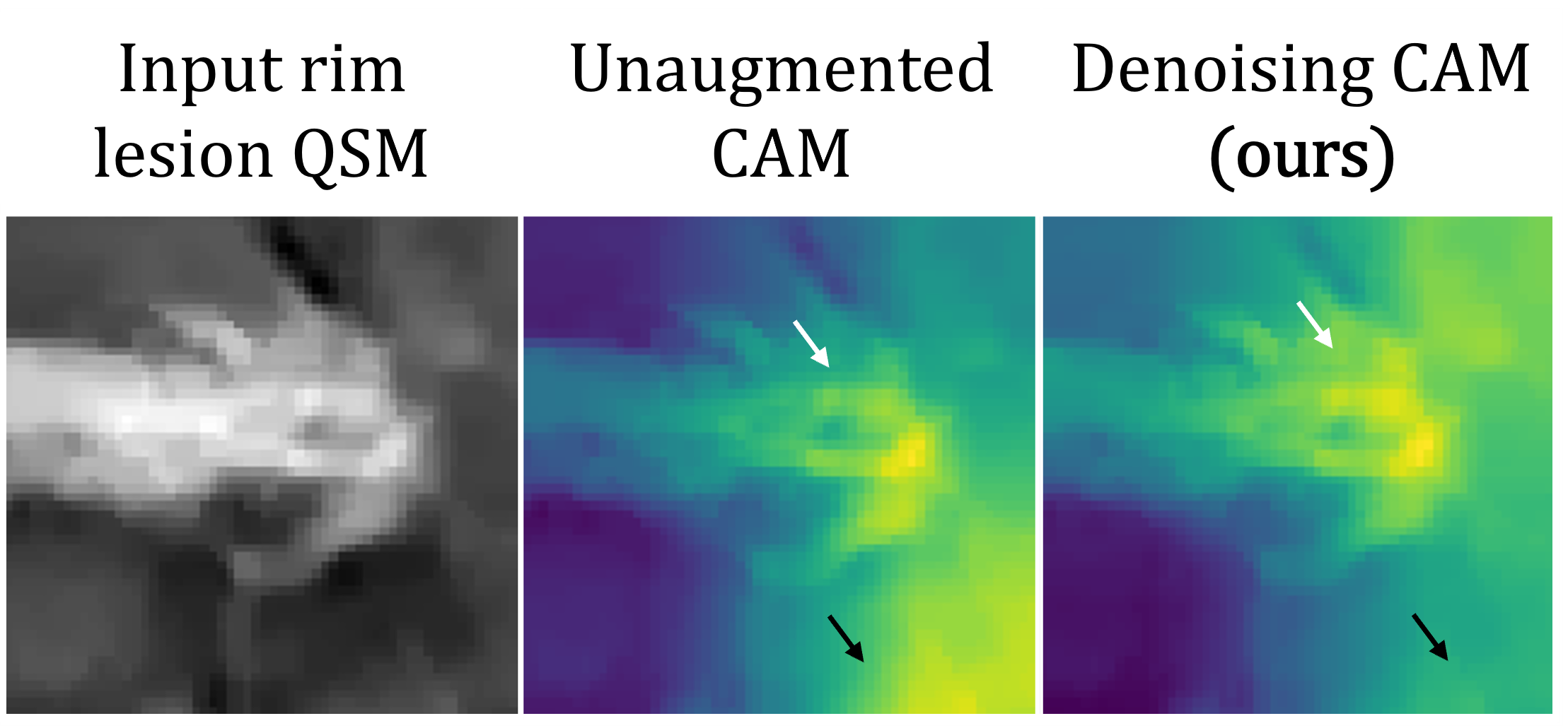}
   \caption{Improved interpretability of class activation maps when using the denoised rim lesion data augmentation, note the shift in intensity from the right corner to the lesion rim (black arrow) and an increase in intensity near the lesion rim (white arrow).}
   \label{f9}
\end{figure}

Both CAMs in Figure \ref{f9} show an emphasis on surrounding white matter, which creates contrast in comparison to demylinated lesions. We notice shift in intensity from the right corner to the lesion rim and note our latent projection denoising augmentation helps identify clinically relevant areas of the lesion. In the map from the classifier trained with our augmentation, both the lesion itself and the hyperintense rim region are highlighted.

\subsection{Multi-contrast extension and segmentation}
\begin{figure}[t]
  \centering
  % \fbox{\rule{0pt}{2in} \rule{0.9\linewidth}{0pt}}
   \includegraphics[width=\linewidth]{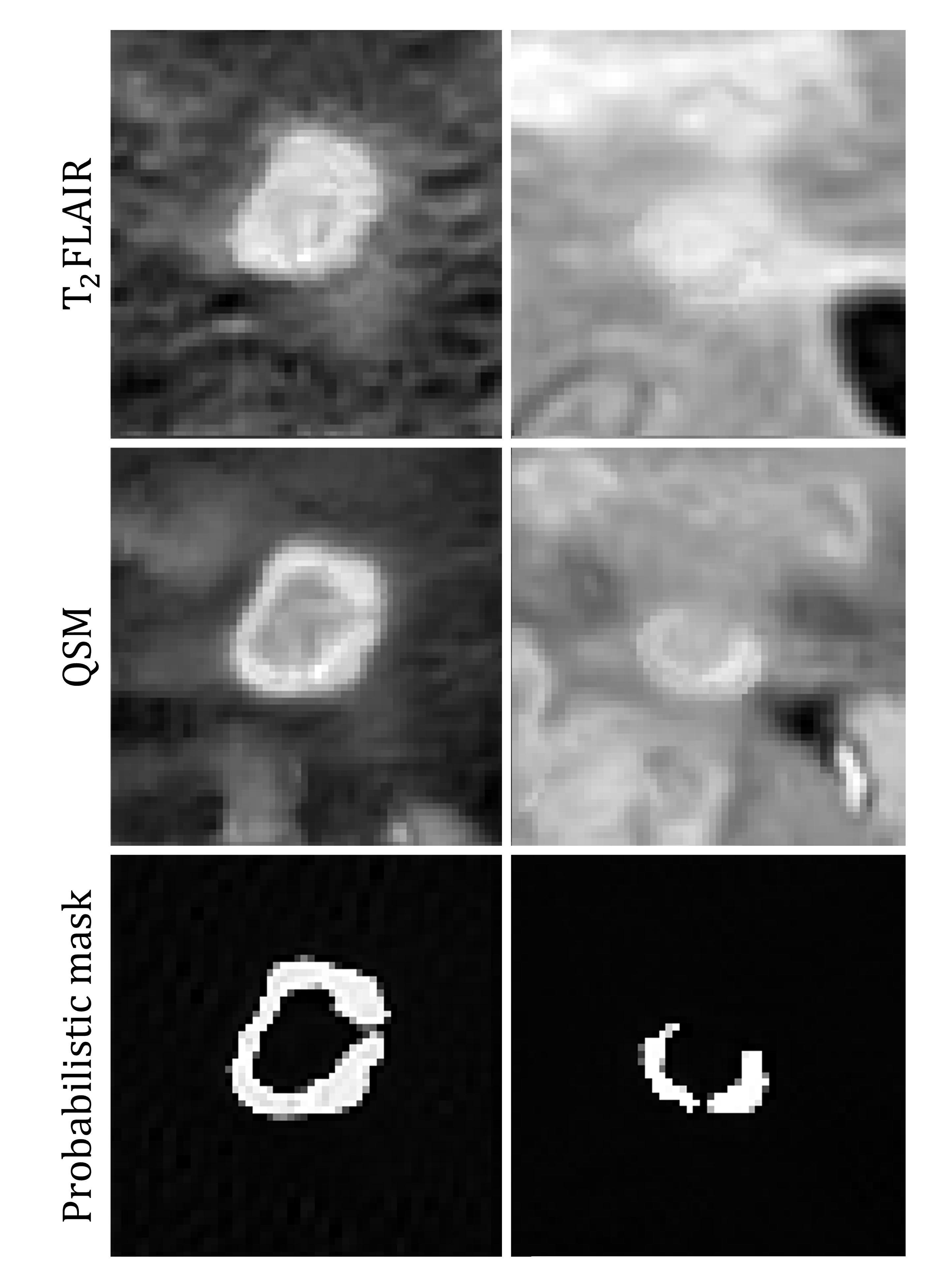}
   \caption{Multi-contrast synthetic example including the lesion susceptibility, probabilistic map, and $\mathrm{T_2FLAIR}$ images.}
   \label{f10}
\end{figure}

Combining the susceptibility map, $\mathrm{T_2FLAIR}$ and paramagnetic rim lesion mask allows realistic generation of new samples as seen in Figure \ref{f10}. We note the realistic lesion depiction across each contrast as $\mathrm{T_2FLAIR}$ lesions typically bound the appearance of the lesion on susceptibility maps. Further, the $\mathrm{T_2FLAIR}$ lesion is uniformly hyperintense, as expected. The generated probabilistic map correctly avoids the hypointense lesion core and correctly identifies the paramagnetic rim. Additionally, a variety of rim lesions are well-represented - both rims encircling around a half of the core circumference (right column) and its entirety (left column) are realistically depicted in both susceptibility map and $\mathrm{T_2FLAIR}$ images with accurate probabilistic segmentation maps.

\newpage
\section{Discussion}
\subsection{Label denoising and class imbalance}
In the medical domain, prognostic tools are commonly marred by some form of rater noise, resulting in class uncertainty~\cite{Jlie} or varied numerical outcomes~\cite{Arob0}. The projection denoising approach we propose only requires access to some unambiguous or noiseless cases to learn a latent space upon which the problematic sample can be projected and increases our minority class dataset by nearly $70\%$. Though this is an active area of research~\cite{Uudd}, the variation in quality of snythesized data can require extensive curation~\cite{Shao}. Our latent projection denoising approach produced an augmented distribution closer to the unseen test rim lesion distribution than the purely or combined synthetic augmentations, which may alleviate curation requirements. Future investigation may focus on latent projection denoising into the conditional latent space, which demonstrated improved convergence during training.

\subsection{Class rebalancing}
We supplement training datasets here with 100 additional lesions for each augmentation method compared but note that the optimal extent of rebalancing with synthetic data is a current topic of interest~\cite{Myeb}. Future investigation may focus on the optimal fraction of synthetic training data for the rim lesion detection problem.

\subsection{Multi-constrast extension}
Jointly generating susceptibility, $\mathrm{T_2FLAIR}$, or other contrasts alongside probabilistic segmentation masks is beneficial beyond the imbalanced class problem. When used in conjunction with augmentation techniques, our method can enable deep learning approaches on small and/or incomplete medical imaging datasets where data collection is slow, laborious and expensive.

\subsection{Applications}
Other susceptibility contrasts such as source separation~\cite{Adim}, myelin imaging~\cite{Msis}, and oxygen extraction fraction~\cite{Msis0} have clinical value in the treatment of multiple sclerosis. We note that extension to these maps using our proposed multi-contrast method is feasible. Our proposed framework can be generalized into higher dimensions to accommodate additional spatial and/or temporal data, and future work should focus on addressing the need~\cite{Xpan,Kden,Echa} for additional training data required for such applications. 

\section{Conclusion}
We examine the quality of synthetic samples via expert radiologist assessment and show that realistic rim lesions can be acquired from generative modeling on multiple contrasts. We demonstrate the effectiveness of our proposed latent projection denoising of ambiguous rim lesions by comparing the FID between different training and unseen test datasets. We further show the improvement in paramagnetic rim lesion detection on QSM with the inclusion of this denoised data and observe increased interpretability of the classifier class activation maps, indicating clinically relevant predictions resulting from augmentation with realistic synthetic data.

{
    \small
    \bibliographystyle{ieeenat_fullname}
    \bibliography{main}
}

% WARNING: do not forget to delete the supplementary pages from your submission 
% \input{sec/X_suppl}

\end{document}